# An experimental investigation of the heat and flow features in street canyons: Impacts of the approaching turbulent boundary layer flow

Yunpeng Xue,[1,a)] Yongling Zhao,[2] Shuo-Jun Mei,[3] Yuan Chao,[4,5] and Jan Carmeliet[2]

[1]*Future Resilient Systems, Singapore-ETH Centre, ETH Zurich, Singapore*
[2]*Department of Mechanical and Process Engineering, ETH Zürich, Zürich, Switzerland*
[3]*School of Atmospheric Sciences, Sun Yat-sen University, Zhuhai, China*
[4]*Department of Architecture, National University of Singapore, Singapore*
[5]*NUS Cities, National University of Singapore, Singapore*

**Abstract**

The study of turbulent boundary layer flow holds significant importance in urban climate research, particularly concerning numerical simulation studies where it serves as a crucial inflow boundary condition. However, understanding the turbulent boundary layer's influence on flow and heat features within canyon and canopy flow remains incomplete. To address this knowledge gap, our current work employs simultaneous Particle Image Velocimetry and Laser-Induced Fluorescence (PIV-LIF) measurements within a large closed-circuit water tunnel. Through this approach, we obtain valuable flow information under various flow and thermal conditions, allowing us to explore the impacts of three distinct turbulent boundary layer flows. The three chosen turbulent boundary layer flows display distinct influences on flow characteristics and heat removal capacity. The ventilation rate exhibits a maximum difference of 80% among the tested boundary layer flows. Additionally, the most significant variation in heat removal capacity is approximately 45%. Moreover, the different turbulence inlet profiles result in diverse fluctuating features at the canyon opening, while the deeper region of the canyon remains less affected.

## I. Introduction

In recent years, summers have been experiencing a concerning trend of increasing heat, leading to heightened attention towards urban climate issues, such as thermal comfort, urban ventilation, and pollutant dispersion in cities. Urban climate studies mostly rely on a simple and widely accepted assumption known as the neutral boundary layer assumption, which neglects thermal stratification in air dynamics. However, in reality, neutral conditions occur relatively rarely during summertime[1]. Urban areas exhibit non-isothermal air flow characteristics, mainly due to the significant temperature difference between the ground or building surfaces and the surrounding air, which can exceed 20-30 degrees Celsius in the urban area[2]. This temperature-driven buoyant force plays a crucial role in shaping the flow structure, air ventilation, heat flux, and temperature distribution within street canyons, especially under calm wind conditions. Thus, comprehending the flow behaviour in urban areas, particularly the buoyant flow in an urban heat island environment, stands as a fundamental and critical aspect of urban climate research[3-5].

Among the various influential factors, the approaching flow condition plays a crucial role in important urban climate impacts, as it directly influences the flow field within the street canyon[6]. Notably, even under neutral stratification, different re-circulating flow behaviours within the street canyon can be observed due to variations in the approaching airflow[7]. Study by Want et al.[8] highlights the profound effect of initial boundary layer conditions on the near wake. Increasing boundary layer thickness enhances the base vortices, resulting in a stronger upwash flow that mitigates the downwash effects of the free-end shear layers. This leads to a decrease in

---

a) Corresponding author. Email: Yunpeng.xue@sec.ethz.ch

Reynolds stresses near the ground and an increase near the free end of the cuboid. Lim et al.[9] reported similar impacts, observing lower Reynolds stresses throughout the entire building height in the near-wake flow region due to the increase in incoming boundary layer thickness. Salizzoni et al.[10, 11] conducted an experimental study where they manipulated the boundary layer condition by adjusting the width of the canyon models upstream of the test canyon. Their findings reveal that the dynamics of the shear layer flow at the top of the canyon and the flow within the cavity are notably affected by turbulent fluxes originating from the approaching boundary layer flow. These observations carry significant implications for establishing parametric relationships governing the turbulent exchange of fluid in the urban canopy.

The presence of a distinct shear layer extending across the cavity top, characterised by shear stress and turbulent kinetic energy (TKE) production, is readily observed when the upstream boundary layer exhibits a laminar or moderately turbulent nature[12]. However, in the case of strong turbulence in the upstream flow, the shear layer's TKE production is not discernible. The measurements indicate significant differences in the shear layer's behaviour when large-scale turbulence and disturbances are present in the upstream flow.

When thermal conditions are introduced into the approaching flow, it introduces a more intricate pattern compared to isothermal flow conditions. Recent experimental findings[13] have unveiled that boundary layer flows with pronounced buoyancy can significantly improve air ventilation and heat removal in specific street canyon configurations. Furthermore, prior reports suggest that in convective stratification scenarios, the friction velocity, a relevant parameter for near-wall dynamics scaling, doesn't accurately represent outer layer behaviour[14]. However, it's been observed that both roughness and unstable stratification contribute to an apparent increase in friction velocity[15].

The impact of turbulent inflow is a crucial consideration in Computational Fluid Dynamics (CFD) studies, as it can lead to varying simulation outcomes in urban climate research. Elevated turbulence intensity in the inflow triggers several noteworthy changes in flow characteristics[16], including heightened turbulent kinetic energy, increased turbulent diffusivity within the street canyon, reinforced street-canyon vortices, and enhanced mean horizontal velocity near the roof level. Consequently, the turbulence intensity of the inflow is found to significantly influence the concentration of pollutants in the street canyon, making it a crucial factor in pollutant dispersion. Similarly, investigations were conducted on the flow characteristics in a cavity with both developing laminar boundary layer and fully turbulent boundary layer flows upstream of the cavity[17]. In cases where fully turbulent inflow and a substantial momentum thickness exist in the incoming boundary layer, the shear layer on top of the cavity exhibits jittering behaviour due to the presence of near-wall coherent structures. These coherent structures strongly impact the formation and convection of eddies within the separated shear layer, leading to dynamic changes in the flow pattern and behaviour of the cavity.

Numerous attempts have been made to accurately simulate the non-equilibrium approaching boundary layer flow. To reduce computational costs, wall functions (WFs) are commonly employed, where the boundary layer flow close to the surface is modelled instead of resolved[3, 18-22]. On the other hand, low-Reynolds number modelling (LRNM) allows for explicit resolution of the boundary layer flow down to the viscous sublayer, offering a more detailed representation[23, 24]. For generating turbulent inflow data in the main domain, the turbulence recycling method is often utilized. This involves adding a subdomain with recycled turbulence upstream of the main domain, along with specially designed buffer regions[25-27]. Another approach used in simulations involves artificially-

a) Corresponding author. Email: Yunpeng.xue@sec.ethz.ch

generated fluctuating boundary layer flow based on experimental data from tunnel approaching flows. However, even with Large Eddy Simulation (LES), discrepancies with experimental data persist in some cases[28-31].

It is essential to acknowledge that the existing methods mentioned above for generating inflow turbulence can only reproduce certain properties, and there are currently no robust techniques available to generate inflow turbulence with all desired characteristics. These characteristics include turbulence intensity, shear stresses, length scales, power spectrum, and coherent turbulent structures spanning various spatial scales down to the Kolmogorov scale, which interact with each other[27, 32-36]. This emphasises the crucial importance of accurately simulating turbulent boundary layer flow in CFD for a wide range of non-isothermal atmospheric studies. One significant challenge arises from unintended changes in streamwise gradients or horizontal inhomogeneity in the vertical profiles of mean wind speed and turbulence quantities as they propagate from the inlet of the computational domain to the modelled buildings. These changes are caused by inconsistencies between the inlet boundary conditions, wall functions, computational grid, and turbulence models, significantly impacting the quality of simulation results[37, 38] and introducing substantial errors[39, 40].

The influence of turbulent inflow, particularly when impacted by thermal-induced buoyant forces, on urban flow mechanisms such as air ventilation and heat transport remains insufficiently understood. Additionally, there is a need to enhance our comprehension of the role of inflow turbulence in improving the accuracy of complex urban flow simulations. Consequently, our current study focuses on investigating the non-isothermal turbulent boundary layer flow within urban environments. To achieve this, we conducted simultaneous Particle Image Velocimetry and Laser-Induced Fluorescence (PIV-LIF) measurements in a large closed-circuit water tunnel. This approach allows us to capture high-resolution data on heat and fluid flow patterns in street canyons under various turbulent layer inflow conditions. By thoroughly analysing the heat and flow behaviours, as well as air ventilation and heat flux characteristics within the street canyons under different approaching turbulent flow and thermal conditions, we gain valuable insights into the impacts of the incoming flow.

## II. Experimental configurations
### a. Experimental facilities and model configuration

This experimental investigation is carried out in the ETH Zurich Atmospheric Boundary Layer Water Tunnel, which is operated at Empa (Swiss Federal Laboratories for Materials Science and Technology). The water tunnel is equipped with a 110-kW pump capable of generating water velocities ranging from 0.02 to 1.5 m/s. Figure 1 (bottom) illustrates its configuration, featuring a 6-meter-long development section and a test section with a cross-sectional area of $0.6 \times 1$ m². To produce three different turbulent boundary layer flows (Fig.1 (1)), a combination of five triangle spires (each with a height of 500 mm) and different sets of rectangular blocks are employed as roughness elements. The dimensions of the two different rectangular blocks are $L \times W \times H = 33 \times 45 \times 24$ and $32 \times 45 \times 99$, respectively. These arrangements are strategically positioned with a longitudinal spacing of 100 mm and a transverse spacing of 70 mm between the two arrays. These roughness elements reduce the flow development section to 5 meter, i.e., the distance between these roughness elements and the tested street canyon.

[a)] Corresponding author. Email: Yunpeng.xue@sec.ethz.ch

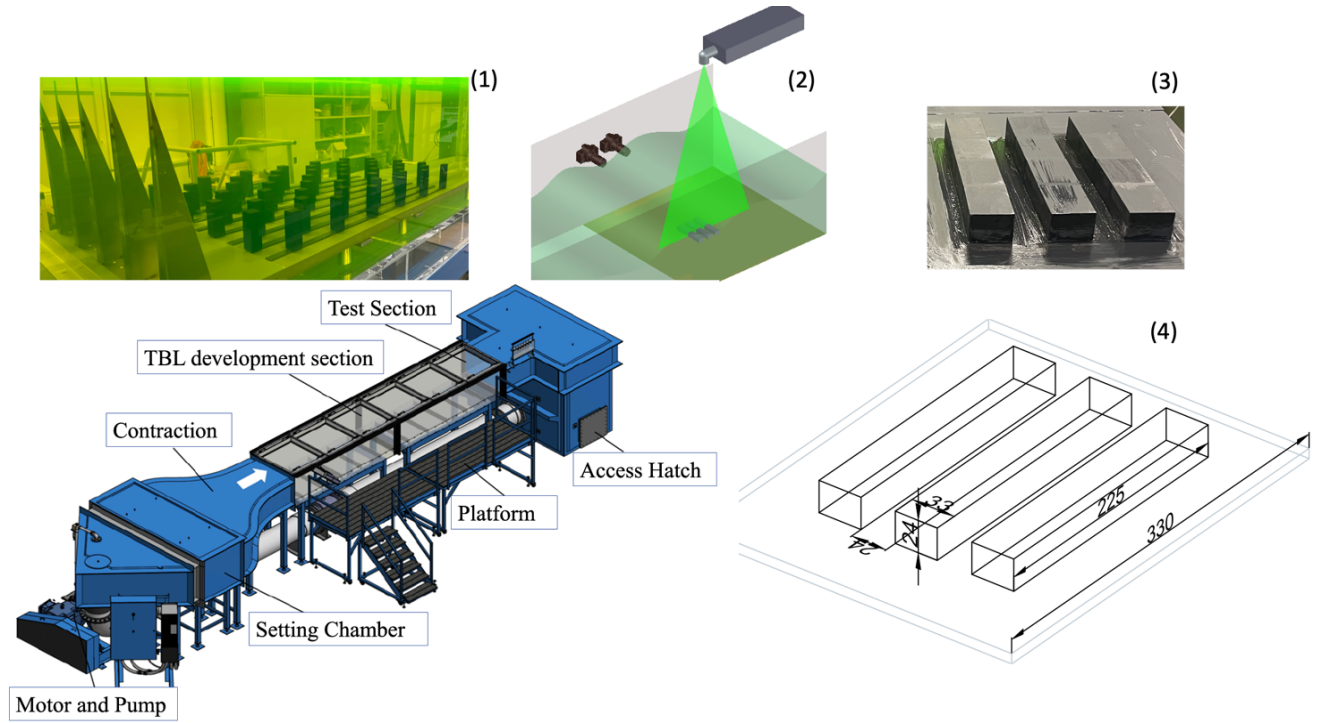

*Figure 1. Illustration of the Atmospheric Boundary Layer Water Tunnel (bottom) and the detailed experimental configuration including: (1) turbulent boundary layer generation elements; (2) PIV-LIF measurement system; (3 & 4) the tested 2D street canyon and model dimensions.*

In this study, a Litron 100 Hz Nd-YAG laser (532 nm) is utilized for dye excitation and flow illumination within the central region of the tested street canyon. The experimental setup comprises two sCMOS 16-bit dual frame cameras, each with a resolution of 2048 × 2048 pixels², operating at 25 Hz. These cameras are aligned in the spanwise direction of the tunnel to focus on the same measurement plane (Fig. 1 (2)). Tracer particles are introduced into the water flow using 10-micron hollow glass beads, resulting in an approximate particle density of 40 to 60 particles within the integration window. For temperature-dependent fluorescence, uranine is employed due to its minimal pulse-to-pulse intensity variation when excited by the laser. Considering that uranine emits fluorescence at a wavelength of 510 nm when excited at 532 nm, fluorescence detection is accomplished using a bandpass filter ranging from 535 nm to 630 nm.

Three conductive models forming a double street canyon are affixed onto a stainless-steel plate, measuring 330 mm × 330 mm, which can be electrically heated to maintain desired temperatures. In this study, one case is set as the isothermal condition ($T_0$) with a temperature of 21°C, while the remaining tested ground temperatures ($T_1 – T_5$) are set at 26 °C, 31 °C, 36 °C, 41 °C, and 45 °C, respectively. These temperatures are continuously monitored using thermal couples attached to the ground surface. The models themselves have dimensions of 33 mm in the longitudinal direction and 225 mm in the spanwise direction. The two street canyons are constructed using building models with a height of 24 mm. The space between the two buildings, or the width of the canyon itself, is also 24 mm, resulting in an aspect ratio of 1 (Fig. 1 (3&4)). Consequently, the street canyon has a length-to-width ratio of 9.375, ensuring a simplified 2-D representation in this research[41]. For further analysis, the non-dimensional height H' is employed and normalised by the model height ($H' = h/H$, where H = 24 mm).

### b. Measurement methods and procedures

[a)] Corresponding author. Email: Yunpeng.xue@sec.ethz.ch

The field-of-view (FOV) for the PIV-LIF measurements is strategically positioned at the centre of the street canyon. This setup effectively captures both the street canyon itself and the region above the building models, allowing for a comprehensive analysis of the in- and outflow dynamics. Velocity fields are acquired through PIV images, which undergo cross-correlation using an integration window of 32 × 32 pixels² (equivalent to 10 pixels/1 mm). To ensure consistency and reliability in our analysis, the average velocity of the upper part of the PIV measurement (located between 180 to 200 mm above the floor) is selected as the free stream velocity for further investigation and analysis in each test case. This chosen reference velocity aids in establishing a consistent basis for our subsequent analysis and findings.

Temperature information is acquired through the post-processing of LIF images. The local intensity of the LIF image is dependent on the concentration of uranine, laser intensity, and fluid temperature. To ensure an accurate linear relationship between the measured local intensity and fluid temperature in a flow with uniformly mixed uranine, we maintain stable laser power and a constant uranine concentration during each measurement. The relationship between local intensity and fluid temperature is calibrated daily, considering the dye's decreasing excitation level over time. Fluctuations in laser intensity, amounting to 2.02%, are determined by analysing images captured from the isothermal case. Additionally, the uranine concentration is consistently maintained at approximately 2 mg/L. However, the transparent box used in calibration introduces some measurement uncertainty, as it adds an extra layer of transparent Perspex, which does not exist during actual measurements. For temperature validation and correction, a series of thermocouples are utilized to measure ground and fluid temperatures at different heights.

Simultaneous measurement of velocity and temperature is accomplished by overlaying the corresponding fields from the same field-of-view (FOV). To achieve reliable statistical analysis, 1500 pairs of images are captured for each test configuration at a frequency of 15 Hz, resulting in a recording time (T) of 100 seconds. The uncertainty of the velocity fields is rigorously estimated to be $10^{-5}$ m/s using integration window-based statistics with sub-pixel accuracy at 1/10 pixel, which is impressively two orders of magnitude smaller than the freestream velocity. Regarding temperature measurements, the uncertainty depends on several factors, including the uranine emission spectrum, optical setup, and pulse-to-pulse laser intensity variation. The intensity-to-temperature ratio is approximately 450, resulting in an instantaneous temperature field uncertainty of 0.002 °C. For time-averaged statistics, the pulse-to-pulse laser intensity variation (2.02%) leads to an uncertainty of approximately 0.09 °C. For readers seeking more comprehensive details on the experimental configurations, calibration, and procedures, we refer them to the comprehensive information available in reference [12].

When describing non-isothermal flow, the bulk Richardson number is used to express the ratio of the buoyancy term to the flow shear term as[13]:

$$Ri = \frac{g\beta \Delta T H}{U_f^2} \qquad (1)$$

where $\Delta T$ is the temperature difference between the floor and the freestream water, $\beta$ denotes the thermal expansion coefficient of water, $g$ is the acceleration due to gravity, $H$ represents the height of the downstream building model, and $U_f$ is the freestream velocity. In Table 1, we present the bulk Richardson number for the water tunnel test conducted in this study. Additionally, Table 1 provides the corresponding Richardson numbers for a

a) Corresponding author. Email: Yunpeng.xue@sec.ethz.ch

full-scale case, using a scaling ratio of 200:1. This full-scale scenario involves a wind speed of 1.7 m/s and maintains an identical temperature difference between the ambient environment and urban surfaces.

Table 1. Richardson numbers for the different test cases and corresponding full-scale case.

| Ground temperature (°C) | | $T_1 = 26$ | $T_2 = 31$ | $T_3 = 36$ | $T_4 = 41$ | $T_5 = 45$ |
|---|---|---|---|---|---|---|
| Freestream velocity (m/s) | $U_1 = 0.03$ | 0.28 | 0.56 | 0.84 | 1.12 | 1.344 |
| | $U_2 = 0.06$ | 0.07 | 0.14 | 0.21 | 0.28 | 0.336 |
| | $U_3 = 0.15$ | 0.011 | 0.022 | 0.034 | 0.045 | 0.054 |
| Full-scale (m/s) | 1.7 | 0.28 | 0.56 | 0.85 | 1.13 | 1.35 |

### c. The turbulent boundary layer flows

In this study, the water velocity in the tunnel is set at approximately 0.03 m/s ($U_1$), 0.06 m/s ($U_2$), and 0.15 m/s ($U_3$), with minor variations (< 1%) due to blockage of the urban model and water evaporation. The first turbulent boundary layer flow (TBL1) is generated solely by the spires, while TBL2 and TBL3 involve a combination of spires and lower or higher roughness elements, respectively. Figure 2 illustrates the profiles of the normalised streamwise velocity distribution, turbulence intensity, and friction velocity at a free stream velocity of 0.03 m/s and approximately 5 H (120 mm) in front of the first model. Comparing TBL1 and TBL2, it can be observed that they have a similar boundary layer thickness ($\delta_{99}$) of about 4.5 H, but TBL2 exhibits slightly stronger turbulence. Additionally, TBL2 also displays a slightly thicker boundary layer than TBL1, as shown in Figure 3 below. On the other hand, TBL3, generated by the combination of spires and larger roughness elements, exhibits both a thicker boundary layer and stronger turbulence. Moreover, the friction velocity reaches its maximum value at about 0.42 H for both TBL1 and TBL2, while TBL3 shows this peak at around 0.57 H.

In Figure 3, the averaged streamwise velocity profiles are compared against the logarithmic universal law of the wall function given by Equation (1):

$$u^+ = \frac{\overline{U}}{U_\tau} = \frac{1}{\kappa}\ln(y^+) + C, \qquad y^+ = \frac{hU_\tau}{\nu} \qquad (1)$$

where $\overline{U}$ represents the averaged streamwise velocity, $U_\tau$ is the friction velocity, $h$ is the height, $\nu$ is the kinematic viscosity of water, and $\kappa$ is the Karman constant. Traditional flat plate values of $C = 5$ and $\kappa = 0.41$ are employed. From the comparison, it is evident that the street canyons, as denoted by the blue dashed line indicating their height, are submerged in the viscous sublayer and buffer layer during the tests, making them significantly influenced by the turbulent boundary layer condition.

[a] Corresponding author. Email: Yunpeng.xue@sec.ethz.ch

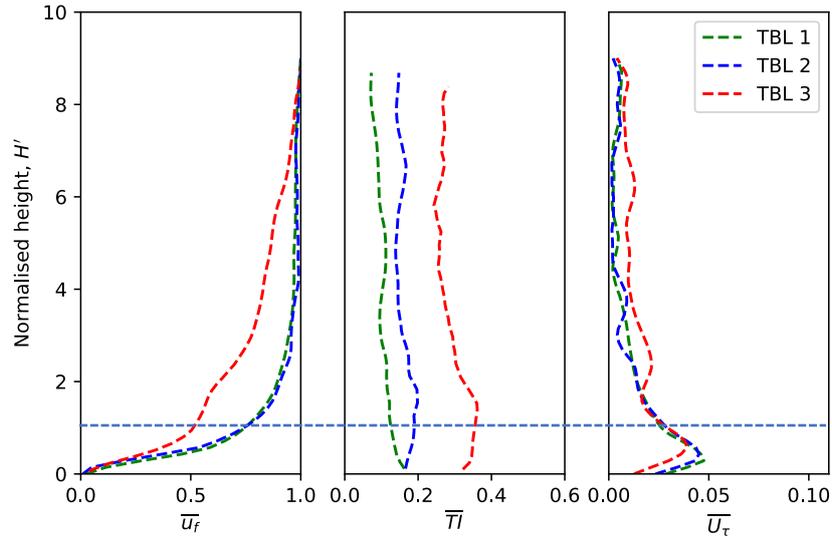

*Figure 2. The profiles of the three turbulent boundary layer flows are presented in the following order from left to right: mean velocity normalised by freestream velocity, turbulence intensity, and normalised friction velocity, all at a freestream velocity of 0.03 m/s.*

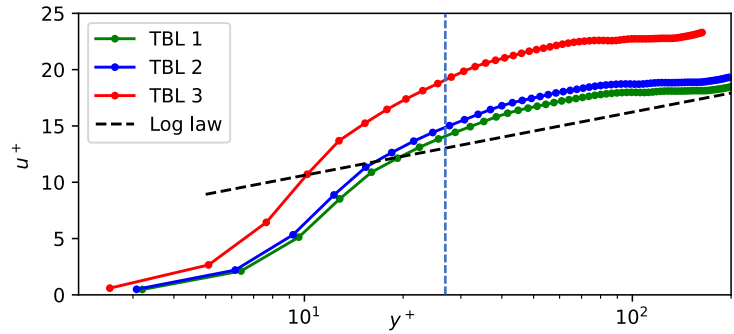

*Figure 3. The average streamwise velocities plotted against the log law of the wall for the three turbulent boundary layer flows.*

### III. Characteristics of the street canyon flow

We initiate our analysis by examining the heat and flow patterns under various conditions, including different ground temperatures and freestream velocities, within the first turbulent boundary layer (TBL1) in this section. Following an example of the temperature and velocity profile (section 3.1), we report the impacts of the thermal condition on the average canyon flow, temperature field and fluctuation characteristics (section 3.2). Subsequently, we compare the effects of these distinct turbulent boundary layers on the properties of canyon flow as well as the heat transfer performance, considering both isothermal ($T_0$) and non-isothermal ($T_4$) cases, in section 3.3.

#### a. Example of the velocity and temperature field

Figure 4 (left) presents an instantaneous temperature field, along with the simultaneous flow pattern, providing a visual representation of the unsteady flow (right to left as indicated by the green arrow) within and around the street canyon, along with the plumes generated by the heated surface. The freestream velocity in this instance is 0.03 m/s, with the fluid temperature on the ground surface measuring approximately 27.5 °C. The obstructed flow over the first model is clearly observed, with distinct thermal plumes developing and shedding from the heated ground and building surfaces downstream. As a result, there is a considerable increase in water temperature of

[a) Corresponding author. Email: Yunpeng.xue@sec.ethz.ch

about 5 to 6 °C within the street canyons. Some of the accumulated heat is carried away by the updraft and subsequently flushed downstream by the approaching flow above the rooftops. The flow pattern is evidently dominated by buoyancy, as indicated by the strong upward flow ascending to about three times the building height. Notably, the heat plumes are observed to be ejected as pulses rather than a continuous flow. These characteristics are smeared out in the average temperature and velocity profiles (Fig. 4 (right)), offering valuable insights into the dominant flow features, the regions significantly impacted by the high-temperature ground, the overall flow behaviour and its relationship with the temperature distribution within and around the street canyon.

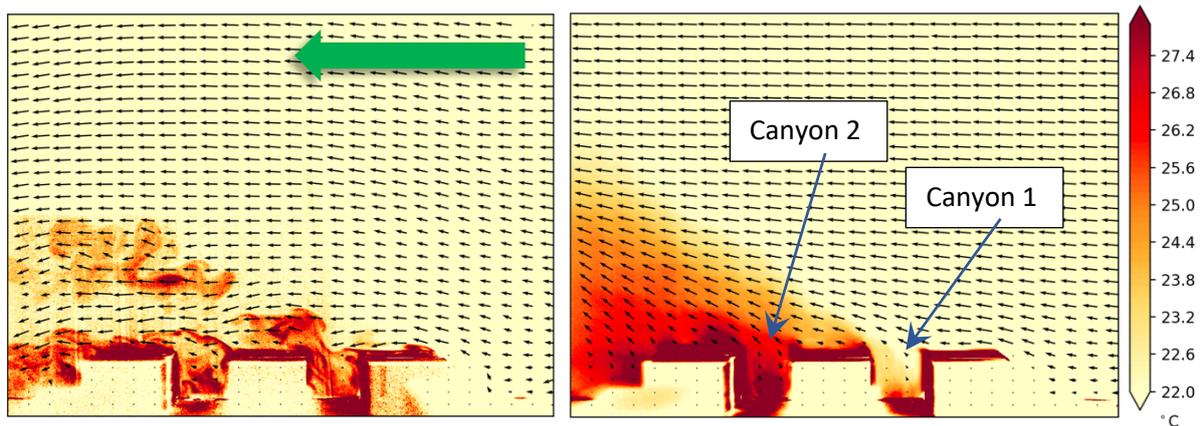

*Figure 4. Normalised instantaneous velocity vector field and the simultaneous temperature field in and around the street canyon (left), and normalised average velocity and temperature profile (right) with a free stream velocity of 0.03 m/s (TBL1) and ground temperature of about 41 degrees.*

### b. The heat and fluid flow characteristics within and over the street canyon

#### i. The different thermal conditions

An increase in ground temperature leads to higher temperatures over a larger region. Figure 5 summarises the average temperature fields for all six temperature cases (with thermal boundary layer TBL1). With the temperature increasing from $T_0$ to $T_5$, the temperature fields exhibit higher temperatures and larger regions heated up. It highlights the significant influence of thermal conditions on the flow dynamics within and around the street canyons. The higher temperatures in the warmer cases correspond to increased buoyant forces, resulting in more pronounced upward motion and stronger thermal plumes.

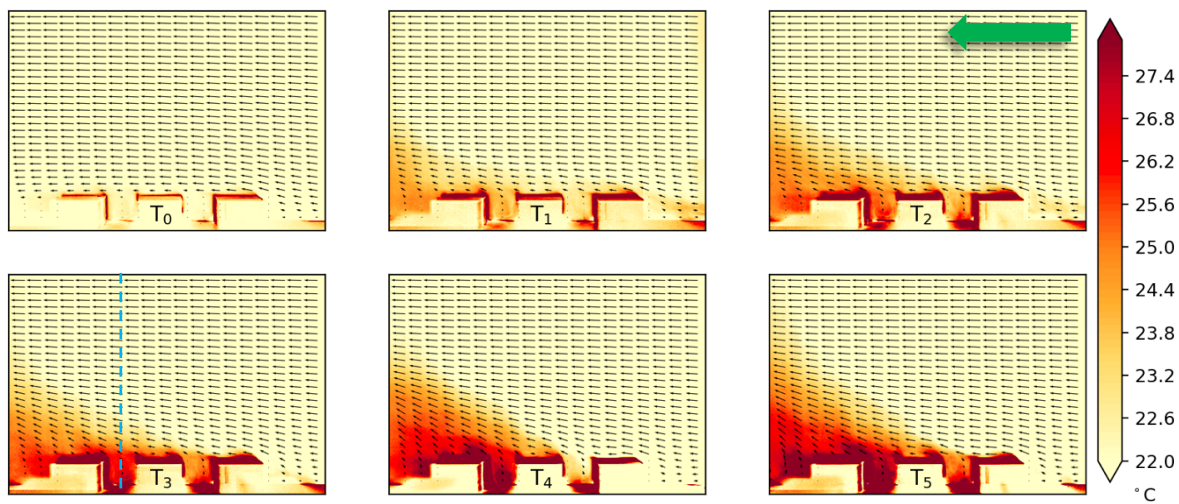

a) Corresponding author. Email: Yunpeng.xue@sec.ethz.ch

*Figure 5. The temperature distribution within and around the street canyons at a freestream velocity of 0.03 m/s (TBL1) and different ground temperatures. The green arrow indicates the flow direction.*

## ii.  Average velocity in different thermal conditions

Figure 6 provides a concise summary of the streamwise velocity, normalised by the freestream velocity of 0.03 m/s (TBL1). The velocity gradient of the flow over the canyons is clearly visible. While a strictly linear relationship between the ground temperature and boundary thickness cannot be conclusively established, it is evident that a significant increase in the ground temperature leads to an expansion of the boundary layer flow. Interestingly, we observe a relatively higher velocity just over the first building model in the non-isothermal tests, as circled for $T_2$, which is caused by the local interaction of the approaching flow and the buoyant updraft flow. Notably, a clear positive correlation between this local higher velocity and the thermal condition can be deduced from the velocity plots. The thickened boundary layer is attributed to the updraft of thermal plumes, which is directly linked to the ground surface temperature as presented in Figure 7.

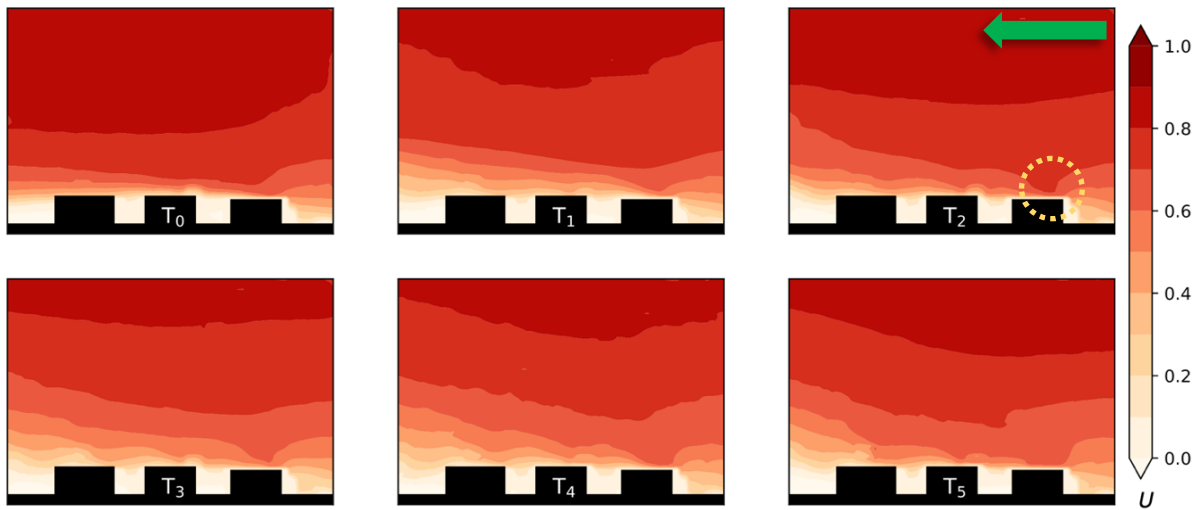

*Figure 6. The streamwise velocity normalised by the freestream velocity of 0.03 m/s (TBL1) at different ground temperatures.*

The vertical components of velocity at different ground temperatures (Fig. 7) highlight the substantial impact of buoyant forces generated from the heated ground. In the isothermal case, various layers of the velocity centred at the front corner of the first building model are visualised, primarily due to the blockage effect on the approaching flow. Beyond the last model, a slight blue region indicates negative vertical velocity, representing a downward flow in the wake. As the ground surface temperature increases, or, in other words, the Richardson number rises, the vertical velocity is significantly influenced, evident by the gradual intensification of the updraft. Unlike the streamwise velocity, it is more apparent in the vertical velocity profiles that the updraft of thermal plumes is directly related to the temperature variations. Moreover, the negative downward flow in the wake observed in the isothermal case disappears entirely in all the heated cases. This demonstrates the significant alteration in flow patterns caused by the presence of thermal plumes and buoyant forces, which suppress downward motion.

a)  Corresponding author. Email:  Yunpeng.xue@sec.ethz.ch

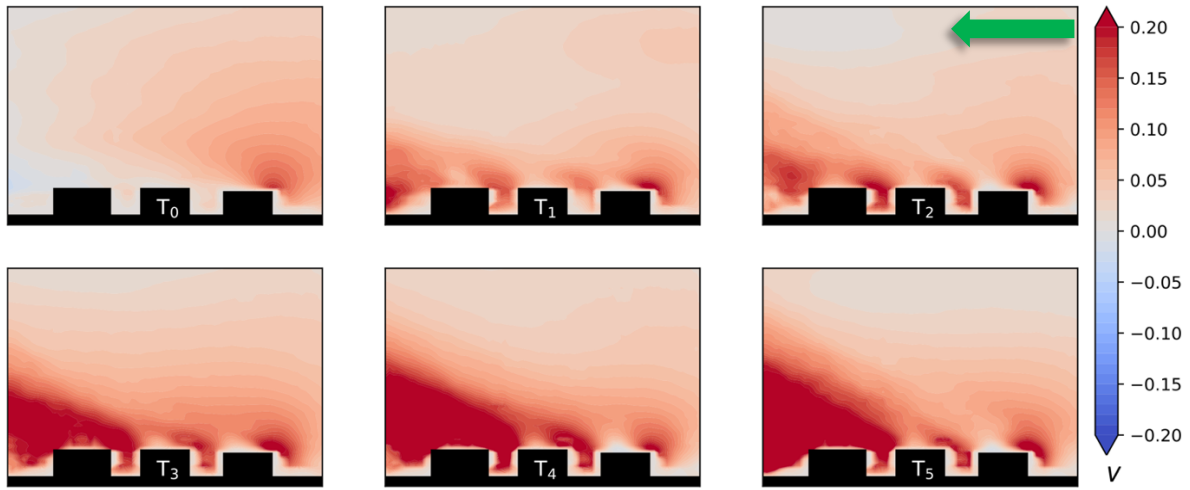

*Figure 7. The vertical velocity normalised by the freestream velocity of 0.03 m/s (TBL1) at different surface temperatures.*

### iii. Impact of freestream velocity

The influence of the freestream velocity is also examined in this study by analysing the normalised vertical velocity for different approaching air velocity conditions, which is directly impacted by the thermal buoyant force. As the freestream velocity increases from 0.03 to 0.15 m/s, similar flow patterns are observed in the isothermal tests, such as the updraft region centred at the corner of the first building block and negative downward flow in the wake (Figure 8). However, a significant difference in the vertical component is evident at the lower freestream velocity ($U_1$) in the non-isothermal case, where the buoyant updraft generated by the high-temperature ground is pronounced. On the other hand, the thermal buoyant force becomes weaker at higher inlet velocity $U_2$ and $U_3$. This is also evidenced by the decrease in the bulk Richardson number from 1.12 to 0.0448 (as shown in Table 1), indicating a substantial weakening impact of the thermal buoyant force as the freestream velocity increases. Indeed, the weakened impact of the thermal buoyant force at higher inlet velocities is consistently observed in all the results. Therefore, we will focus solely on the heat and flow characteristics at the lowest freestream velocity of 0.03 m/s, $U_1$, in the following sections, where the thermal buoyant force has a more dominant influence.

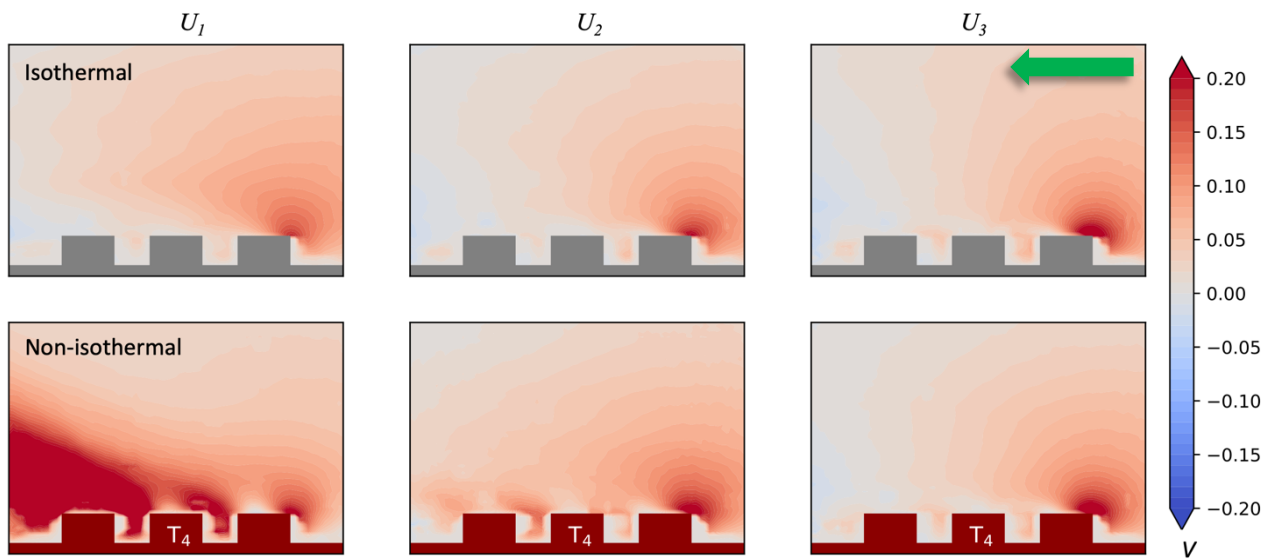

a) Corresponding author. Email: Yunpeng.xue@sec.ethz.ch

Figure 8. The normalised vertical velocity at the freestream velocity of 0.03 m/s, 0.06 m/s and 0.15 m/s ($U_1$, $U_2$ & $U_3$, TBL1), under isothermal (grey models) and heating conditions (red models).

### iv. The fluctuating canyon ventilation

The dimensionless volumetric ventilation rate, $Q'$, from street canyons is another important parameter used to evaluate the urban climate performance, which is calculated based on the instantaneous vertical velocity ($v$) at the canyon roof level, given as[13]:

$$Q' = \frac{\tau}{V_C} \int_A v \, dA \qquad (2)$$

where $\tau$ denotes the reference time and is calculated as $\tau = 2(H + W)/(2U_f/3)$, $V_C$ represents the unit air volume of the canyon, and $A$ is the ventilation area of interest. Figure 9 summarises the ventilation rate at the roof level of the second street canyon as function of time for different ground surface temperatures, illustrating the profound impact of thermal conditions on upward buoyant flow and ventilation. In the isothermal condition ($T_0$), the small magnitude of both positive and negative ventilation rate indicates a relatively weak interaction between the canyon flow and the overlying flow. As the ground surface temperature increases ($T_1$ to $T_5$), the overall ventilation shows a clear trend of strengthening. The positive ventilation indicates a slow outward flow of fluid within the street canyon, and this tendency becomes notably enhanced with higher ground temperatures. The figure demonstrates that an increase in ground temperature leads to a more robust upward buoyant flow and, consequently, greater ventilation as expected here and observed in many studies. Moreover, the increasing frequency of stripes, representing peak ventilation rates due to the plumes, implies a greater fluctuating feature at higher temperatures. This suggests a higher level of temporal variability in ventilation as the ground surface temperature rises.

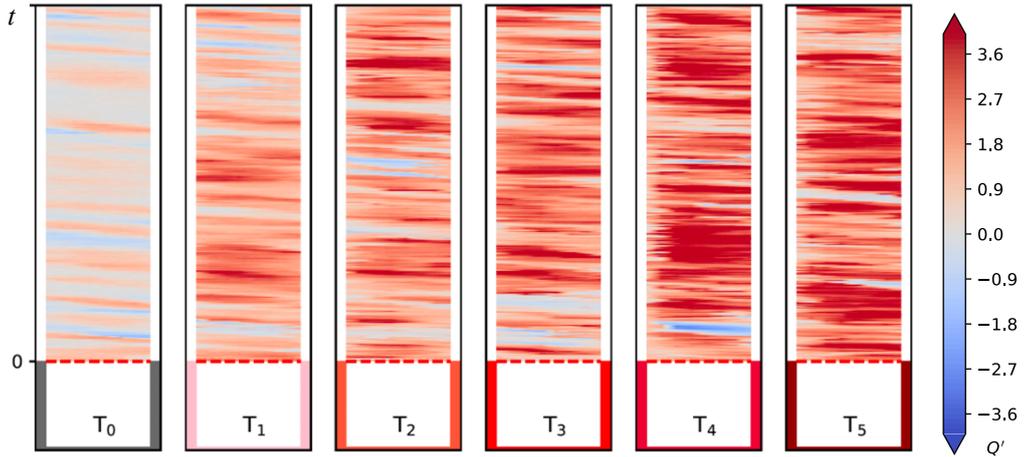

Figure 9. The temporospatial plot of the ventilation rates at the second canyon roof level for the different ground temperatures ($U_1$, TBL1).

### v. The turbulent structure

The unstable nature of the flow can be characterised using second-order statistics of the velocity components. Quadrant analysis, a valuable tool, helps to identify the dominant modes of momentum transport and differentiate between various turbulent transport processes occurring in street canyon flow. By analysing the streamwise (u′) and vertical (v′) fluctuations of the velocity components, we can characterise the momentum flux (u′ × v′) as an

a) Corresponding author. Email: Yunpeng.xue@sec.ethz.ch

outward interaction (u′ > 0, v′ > 0), ejection (u′ < 0, v′ > 0), inward interaction (u′ < 0, v′ < 0), or sweep (u′ > 0, v′ < 0). In this study, we found that the inward and outward interactions have negligible impact, suggesting that the dominant events in all cases are ejections and sweeps.

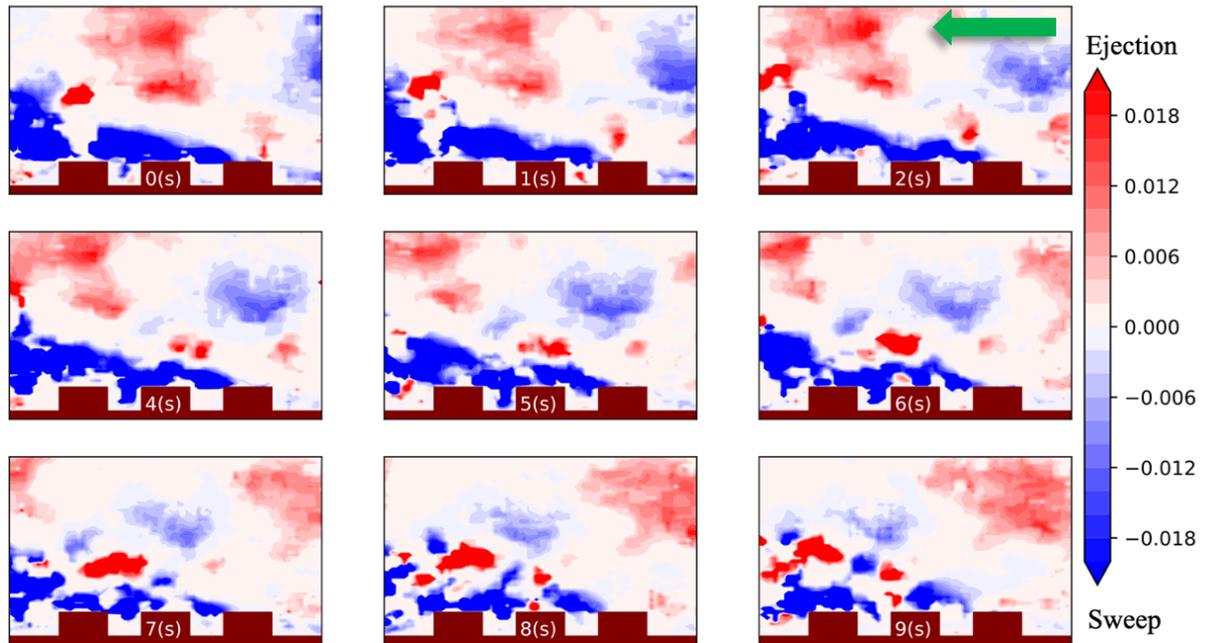

Figure 10. Example of the temporal ejections (red) & sweeps (blue) sampled at 1 Hz ($U_1$, TBL1, $T_4$).

Figure 10 presents an example of the time-series quadrant data for ejections (shown in red) and sweeps (shown in blue). These data are sampled at 1 Hz and normalised by the freestream velocity ($U_f^2$). The strong ejections and sweeps depicted in the temporal series subplots clearly illustrate the unsteady plumes and turbulent structures of fluid being flushed downwards by the approaching flow, and being ejected meanwhile. The time-series plots also reveal the dissipation and generation of these events, implying the unsteady and fluctuating nature of the non-isothermal flow within and above the canyons. Notably, while the current plot primarily shows sweeps near the canyon, examination of other time-series plots indicates that ejections are also observed in the lower region. This further emphasises the significant fluctuating mechanism of the buoyant-driven flow.

###    vi.    The fluctuation of centerline temperature

Thanks to the simultaneous PIV/LIF technique, we can obtain high-resolution temperature fluctuation data along the centreline of the second canyon as depicted in Figure 5 ($T_3$), and subsequently plotted versus time in Figure 11. It's important to note that at the measurement plane, the ground is situated at H′ = 0, and that H′ < 0 refers to the area of ground located in front of the measurement plane captured in the images. From the temperature fluctuation plot, it is evident that most of the heated fluid is concentrated within a region of about 1-2 H (canyon height), indicating the strong thermal impacts within the canyon that gradually decay at higher levels. The presence of flame-shaped high-temperature fluid is observed, extending as high as 3.8 to 3.9 H. This indicates that the influence of the high-temperature ground has a limited impact on the fluid beyond approximately 4 H, at the freestream velocity of 0.03 m/s. In this temporospatial plot, the continuous trace within the figure unfolds the temporal sequence, offering insights into the dynamic process of heat transfer. For instance, the steadily intensifying high-temperature fluid, illustrated by the dark red trace marked along the green dashed line, depicts the upward movement of heated fluid. This phenomenon, akin to updraft heat plumes observed in video records

a) Corresponding author. Email: Yunpeng.xue@sec.ethz.ch

within the street canyon, is evident. The variations and fluctuations in these continuous high-temperature traces and their slopes reflect the unsteady and turbulent characteristics inherent to these heat plumes.

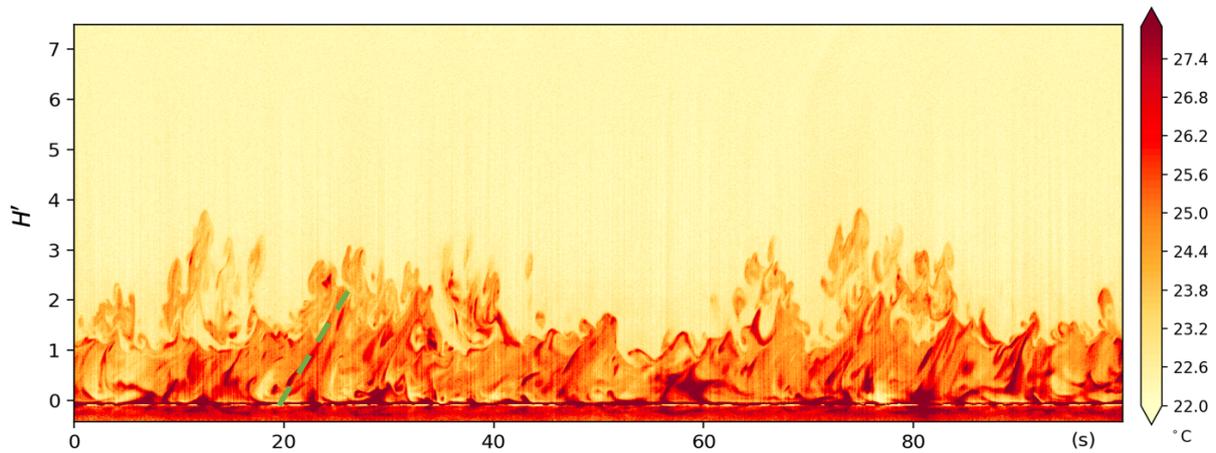

*Figure 11. A temporospatial plot of the temperature along the centreline of the second street canyon as shown in Fig. 5 at a freestream velocity of 0.03 m/s (TBL1) and ground surface temperature of 36 °C.*

### c. Impacts of the different turbulent boundary layers
#### i. Impact on the average velocity

Figure 12 presents the summary of streamwise velocities for the three different approaching turbulent boundary layer flows, listed in two rows: isothermal cases (grey models) and non-isothermal cases (T$_4$, red models, with a ground temperature of 41 °C). In the isothermal tests, the velocity profiles show good agreement with the inlet boundary layer conditions, with similar boundary gradients in TBL1 & TBL2 and a thicker boundary layer in TBL3. However, with the introduction of a heated surface, an obvious thickening of the boundary layer is induced by the updraft buoyant flow in TBL1 & TBL2. On the other hand, the thermal impact on the thick boundary layer (TBL3) is not as significant as in TBL1 and TBL2, primarily due to the smaller difference between the streamwise velocity of the updraft buoyant flow and the overlaying flow. Across all three approaching flows with different turbulent boundary layers, the velocity over the first canyon model increases due to the higher temperature of the top surface in the heating conditions. Therefore, in terms of streamwise velocity, the thermal buoyant force has a more significant impact on the thinner approaching boundary layer flow, owing to the relatively higher velocity at the street canyon opening.

---

[a)] Corresponding author. Email: Yunpeng.xue@sec.ethz.ch

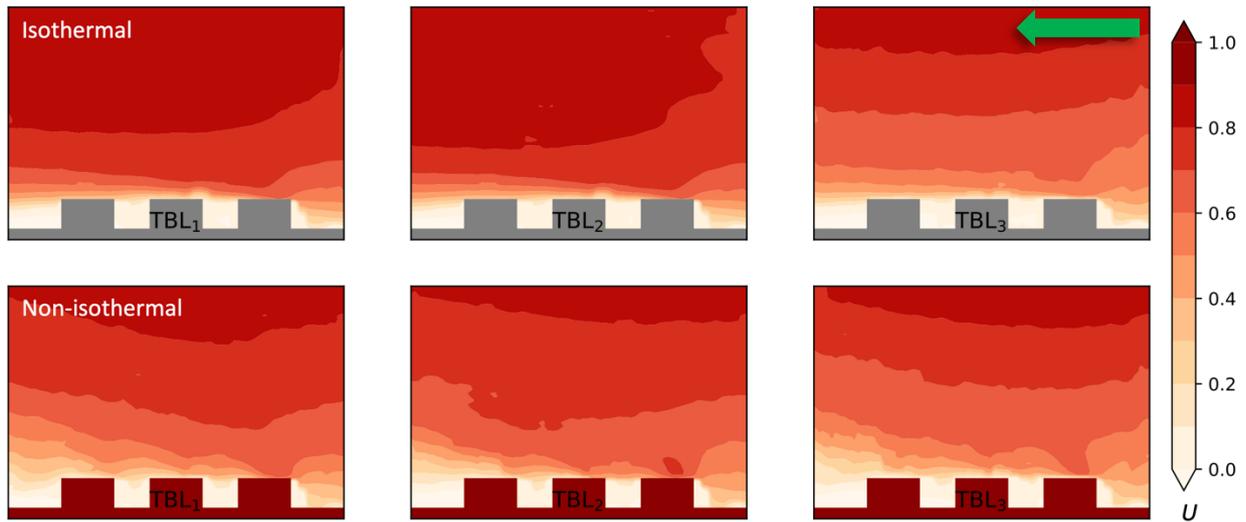

*Figure 12. The normalised streamwise velocity for different boundary layer flows and thermal conditions ($U_1$, $T_0$ & $T_4$).*

Figure 13 compares the vertical component of velocity, which is significantly influenced by the thermal buoyant force in the non-isothermal cases, for the three tested turbulent boundary layer inflows. In the isothermal case, the thicker boundary layer (TBL3) exhibits relatively lower velocity in the region close to the ground, resulting in a more pronounced negative downward flow in the wake of the models. Under the non-isothermal conditions, all three boundary layer flows show a clear increase in upward buoyant flow within and after the heated canyon area. TBL1 appears to lead to a greater increase in vertical velocity in the non-isothermal tests, while TBL2 shows less significance, particularly as indicated in the first street canyon. However, it is challenging to draw any significant difference from the summarised plots of the vertical velocity generated by the different turbulent boundary layer flows. Therefore, further quantitative analysis of the thermal buoyant force is necessary to gain a better understanding of its influence on the flow dynamics within the street canyon. These findings contribute to unravelling the intricate interactions between thermal buoyancy and turbulent boundary layer flows, enhancing our comprehension of the complex urban climate and its fluid dynamics.

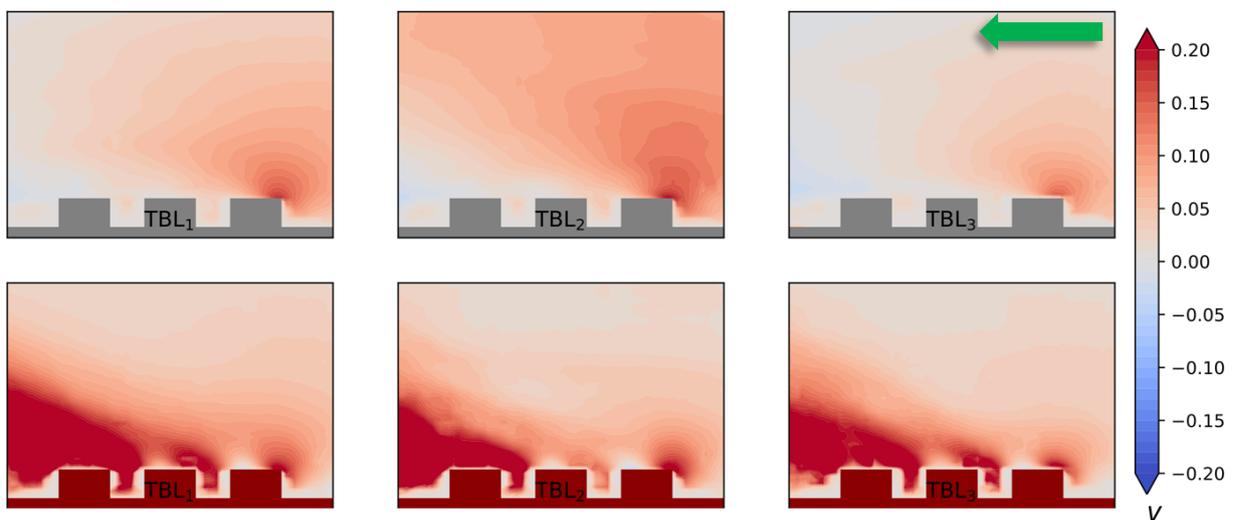

*Figure 13. The normalised vertical velocity in different boundary layer flows and thermal conditions ($U_1$, $T_0$ & $T_4$).*

### ii. Impact on the turbulent intensity

a) Corresponding author. Email: Yunpeng.xue@sec.ethz.ch

The varying TI levels in different turbulent boundary layer flows have significant implications for the dispersion and mixing of pollutants, heat transfer, and ventilation patterns within urban environments. Figure 14 examines the fluctuating components of velocity using time series measurement data and presents the freestream turbulent intensity (TI) in both non-isothermal and isothermal cases. In the isothermal cases, there is a slight increase in the TI from TBL1 to TBL3, visualizing the difference in turbulent intensity in the approaching boundary layer flow. Since the fluctuation is normalised by the freestream velocity in this study, the lower magnitude of velocity in the street canyons, compared to the freestream velocity, leads to lower turbulent intensity in the canyons under isothermal conditions.

On the other hand, in the three non-isothermal cases, increases in freestream turbulent intensity within the canyons are observed, which can lead to enhanced mixing within them. The buoyant updraft induced by the heating source also contributes to a significant increase in the TI over and after the canyons. Moreover, the higher TI in TBL2 and TBL3 allows the heating source to impact a larger region. The stronger fluctuation in these cases is capable of transferring and spreading the induced heat to a larger area, leading to increased local turbulence and a larger region with an obvious increase in TI.

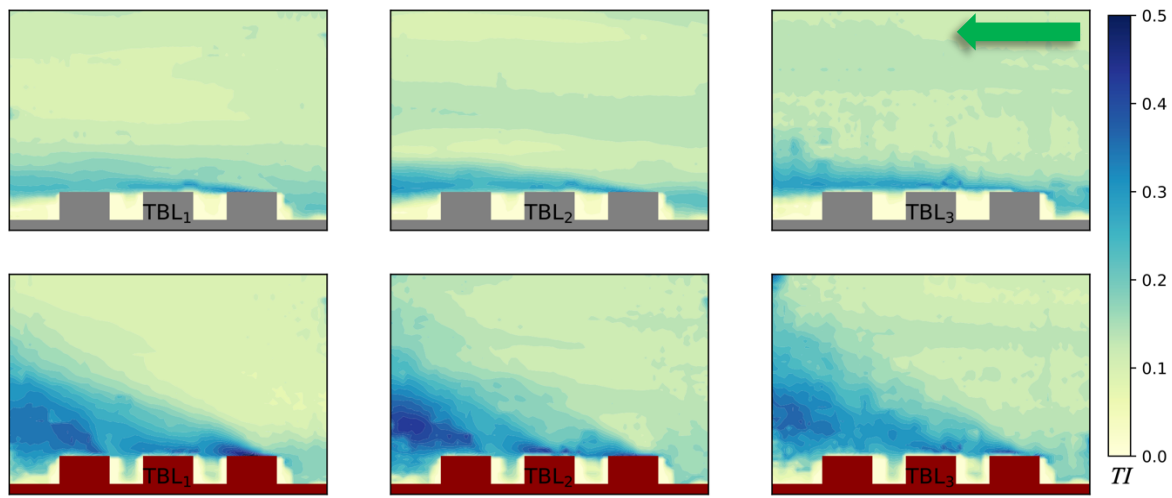

*Figure 14. The Turbulent intensity profiles at different inlet boundary layer and thermal conditions ($U_1$, $T_0$ & $T_4$).*

### iii. Impact on the canyon ventilation

Figure 15 presents an example of the time series of canyon-wise ventilation rates by integrating the ventilation over the canyon width (Fig. 9). The grey and red lines represent the fluctuating ventilation rates for isothermal ($T_0$) and non-isothermal heating ($T_4$) cases, respectively, with the dashed lines indicating the averaged values. When the ground is heated to a high temperature, the buoyancy force generates an updraft flow, leading to a significant increase in the ventilation rate. Moreover, strong fluctuations in the heating situation are observed, reflecting the unsteady updraft flow driven by the buoyancy force or heat plumes. For instance, when the approaching flow is set at 0.03 m/s, the average ventilation rate increases from 0.179 in the isothermal condition to 2.236 at a ground temperature of around 41 °C.

[a] Corresponding author. Email: Yunpeng.xue@sec.ethz.ch

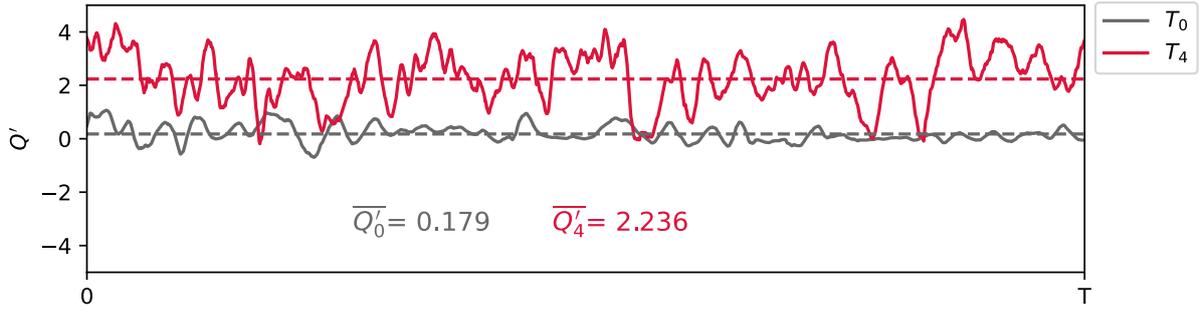

*Figure 15. Time series of the ventilation rate at the canyon opening in isothermal (grey) and non-isothermal heating (red) conditions with the averaged values given at the bottom (TBL1).*

These average ventilation rates under different boundary flow and thermal conditions are then summarised in Figure 16, with the buoyancy-induced increase marked as '×'. Notably, a consistent observation is that the ventilation rates in the first canyon are always smaller than those in the second canyon, regardless of the same thermal conditions within the street canyons. This difference in ventilation is primarily attributed to the buoyancy force of the approaching flow toward the canyons. In the case of the first canyon, the approaching flow is isothermal, and this results in a more significant suppression effect on the ventilation rates. However, once the flow passes the first canyon, it experiences an increase in temperature due to the gained updraft buoyancy, leading to higher ventilation rates in the second canyon. This phenomenon, which we previously discussed in our earlier work[12], demonstrates the suppression effect of the buoyant approaching flow on ventilation rates.

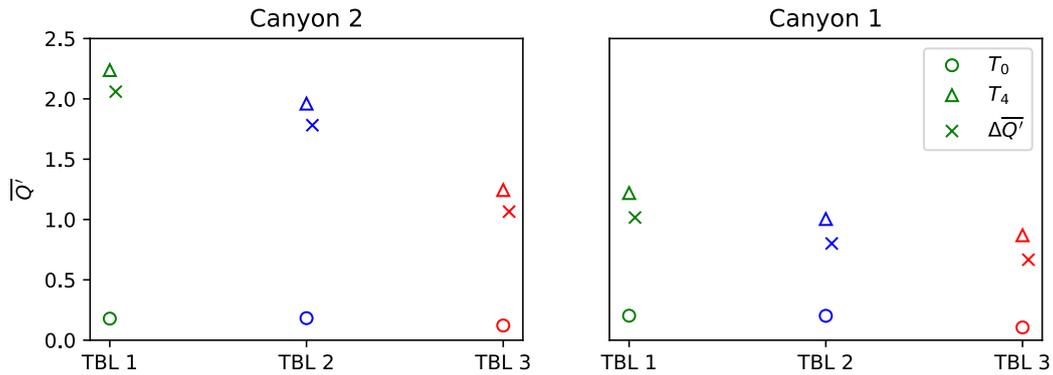

*Figure 16. Averaged ventilation rates at the canyon opening in different boundary and thermal conditions ($U_1$, $T_0$ & $T_4$).*

In both the first and second street canyons, under isothermal conditions, similar ventilation rates are observed. However, in the non-isothermal heating scenario, TBL1 results in the strongest ventilation rate, followed by TBL2 and TBL3, with ventilation rates of 2.24, 1.96, and 1.24, respectively. A substantial increase of up to 80% in the ventilation rate is observed from TBL3 to TBL1.

The differences in ventilation rates among the three turbulent boundary layer inflows can be mainly attributed to their distinct velocity profiles at the canyon roof level. In TBL1 and TBL2, the higher velocity or thinner shear layer at the canyon opening allows more plumes from the ground to penetrate the shear layer flow, resulting in a larger ventilation rate. This means that an increase in the velocity at the canyon opening can lead to more ventilating flow from the canyon, and due to the same freestream velocity or reference time used in this work, the increased ventilating flow is reflected by the larger ventilation rate. However, it is essential to consider that further increases in the velocity may cause a stronger suppression effect on the updraft buoyant flow, resulting in weaker ventilation flow or ventilation rate. This observation may seem contrary to our recent report[42], where higher freestream velocities are associated with smaller ventilation rates. However, it is important to note that an increase in the freestream velocity also leads to a decrease in the reference time, which can subsequently result in a smaller ventilation rate even with an enhanced ventilating flow. To clarify the relationship between the freestream velocity and flow ventilation, a comprehensive analysis of the ventilating flow with or without normalisation by reference time is recommended.

[a] Corresponding author. Email: Yunpeng.xue@sec.ethz.ch

The variation in ventilation rates observed between TBL1 and TBL2 may be attributed to the slight differences in the boundary layer profiles. Additionally, the stronger fluctuations observed in TBL3 may suppress outflow and limit ventilation due to increased mixing. Nevertheless, at this stage, there is insufficient evidence to definitively distinguish between the impacts of the mean and fluctuating components on ventilation rates. Further investigation and analysis are necessary to shed light on this aspect.

### iv. Impact on the turbulence structure

In Figure 17, the accumulated intensities of the four types of events from the quadrant analysis are integrated over time ($\int u'v' dt$) for different non-isothermal boundary layer flows ($T_4$). The focus is on the canyon centreline at different heights from the ground ($H_0$ to $H_3$ with a 6 mm step). It is important to note that the strength of the sweeps and ejections depends on the location of the observation point. Within the street canyon ($H_0$-$H_2$), the fluctuating products have small magnitudes, while at the centre of the canyon opening ($H_3$), they exhibit more significant values. This observation confirms that the sweeps and ejections are much more dominant than the outward and inward interactions in all cases, as mentioned in section 3.2.5. Furthermore, for the sake of clarity and comparison, we have included integrations at $H_3$ for isothermal flows, denoted as $H_3$-Iso. This addition starkly underscores the substantial influence of thermal conditions on these fluctuating characteristics.

It is notable that, despite having nearly identical average velocity profiles, TBL2 exhibits more significant ejections and sweeps compared to TBL1 due to its higher turbulent intensity. Conversely, while TBL3 possesses higher turbulent intensity, the integration magnitude of ejections and sweeps is not as pronounced as in TBL2, primarily due to the slower shear layer flow at the canyon opening. Another noteworthy observation is the relative discrepancy in final values between ejections and sweeps. Specifically, for TBL1 and TBL2, the final ejection values at $H_3$ are smaller than the final sweep values, whereas in TBL3, this pattern is reversed. Although the final value for TBL3 is smaller than for TBL1 and TBL2, it promotes relatively more ejections than sweeps. This observation may offer insights into the differences in ventilation rates observed between these turbulent boundary layer flows.

[a] Corresponding author. Email: Yunpeng.xue@sec.ethz.ch

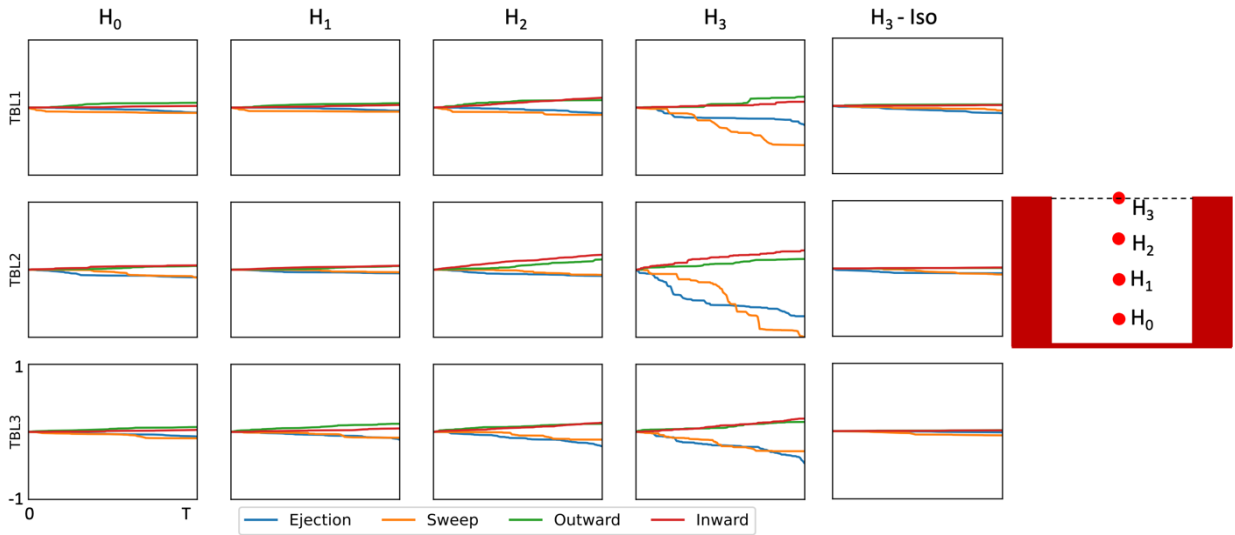

*Figure 17. Time integrations of four events in quadrant analysis ($\int u'v'dt$) at various heights in the canyon ($H_0$-$H_3$) in different non-isothermal turbulent boundary layer flows ($T_4$). Isothermal cases ($H_3$-Iso) are included for enhanced comparisons.*

### v. Impact on the centerline temperature

Figure 18 illustrates the temperature distributions along the centreline in the two street canyons under different turbulent boundary layers and thermal conditions ($T_0$ & $T_4$). The red dashed line represents the canyon centreline, along which the fluid temperature is plotted, the black dashed line indicates the roof level of the canyon, and the vertical yellow dash line indicates the ambient temperature of the flow or temperature of the isothermal flow ($T_0$). It is reasonable to observe similar temperature distributions along the canyon centreline due to the only difference in the approaching flow. However, some zigzag shapes in the temperature distribution, particularly in the first canyon, are observed in all considered conditions.

Upon closer inspection, some differences are noticeable due to the different approaching flow. Over the first canyon in TBL3, the flow temperature is slightly higher than in the other two cases due to the slower flow overlaying the model, allowing for a longer time for the heat transfer process from the surface to the fluid without being removed. At a further higher location, approximately 3.1 H, a weaker updraft flow is evident, as indicated by the vertical velocity profiles in Fig. 13, as well as the smallest ventilation results of TBL3 in Fig. 16. Additionally, the strong fluctuating component in TBL3 enhances the turbulent heat transfer, causing the flow to reach the steady ambient temperature earlier at a lower location. As the flow moves downstream to the second canyon, the updraft buoyant flow with higher temperature leads to the steady ambient temperature being reached at a higher location, approximately 4.5 H.

[a)] Corresponding author. Email: Yunpeng.xue@sec.ethz.ch

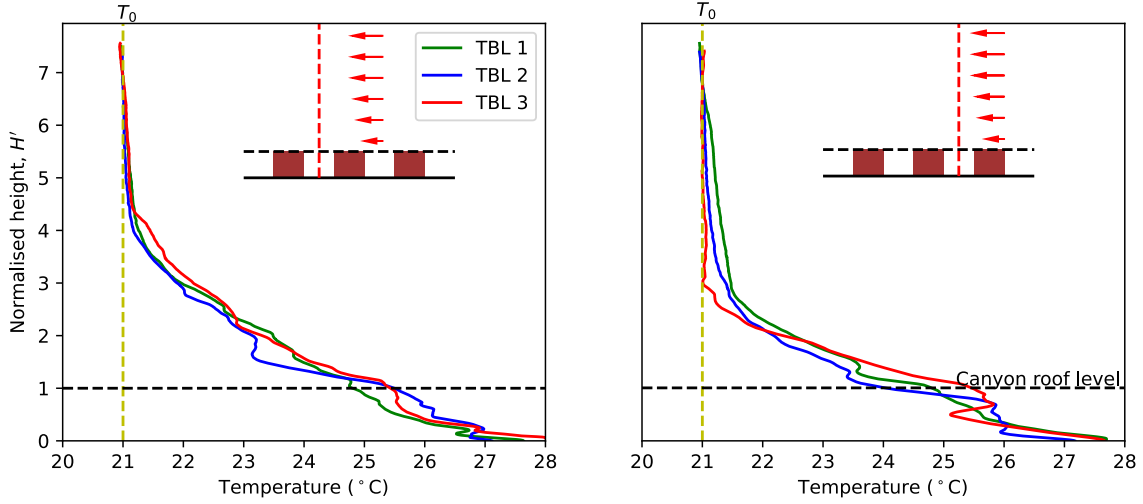

*Figure 18. The temperature distributions along the canyon centrelines (shown as red dashed lines) in different boundary conditions. The black dashed line indicates the roof level of the canyon and the yellow dashed line at 21 degrees indicates the ambient temperature of the flow.*

### vi. Impact on the heat removal performance

In the non-isothermal flow, the heat removal capacity from the street canyon becomes an important measure. The buoyant flow significantly contributes to heat removal from the canyon by providing additional upward momentum and inducing a vertical heat flux from the ground to the free stream flow. With the detailed velocity and temperature profiles, the normalised convective heat flux at the roof level can be estimated as follows[43]:

$$\varnothing' = \overline{V}(\overline{T_r} - T_f)c_p \rho dA / U_f T_f c_p \rho dA \tag{3}$$

where $\overline{V}$ and $\overline{T_r}$ are the roof level time-averaged vertical velocity and fluid temperature, $T_f$ is the freestream temperature, $c_p$ is the heat capacity and $\rho$ is the fluid density. Figure 19 illustrates the heat flux from the canyon openings in different turbulent boundary layer flows. Similar to the ventilation results shown in Figure 15, the heat fluxes from the second canyon opening are greater than those from the first canyon due to the weaker suppression effect of the buoyancy-driven approaching flow. TBL1 and TBL2 exhibit similar profiles because of their similar boundary layers. In contrast, TBL3 results in a lower heat flux as it has a thicker boundary layer, indicating slower shear layer flow at the canyon opening. For instance, the averaged heat flux from the second canyon is 0.11, 0.104, and 0.076 in the three TBLs, respectively. Consequently, a 45% increase in heat flux can be achieved when TBL3 is changed to TBL1 in the second canyon. Noticeable peak values near the downstream edge of the canyon are observed in both canyons and all three TBLs. These peaks are caused by the higher temperature near the model surface and higher vertical velocity near the windward wall of the canyon. A notable contrast of 14.3% is evident in the ventilation rates between TBL1 and TBL2, as depicted in Figure 15. In comparison, the disparity in heat flux between TBL1 and TBL2 is relatively smaller, around 5.7%. This small increase in the ventilation rate could potentially be attributed to the distinct levels of turbulent intensity characterizing TBL1 and TBL2. Such variability in turbulence may exert a discernible influence on the heat flux by impacting turbulent heat transfer mechanisms. Hence, a more comprehensive exploration involving flow scenarios with similar average flow attributes yet varying turbulence levels could offer valuable insights into disentangling the contributions of convective heat transfer and turbulent heat transfer within the context of non-isothermal flows.

[a] Corresponding author. Email: Yunpeng.xue@sec.ethz.ch

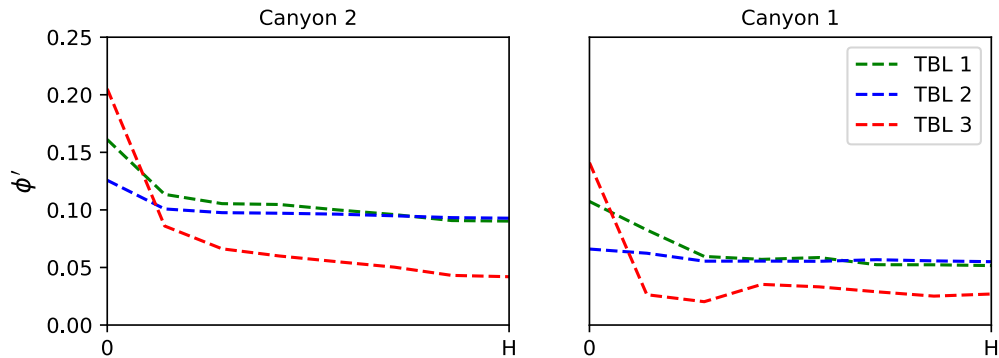

Figure 19. Normalised heat flux at the canyon opening of all cases over the street canyon width ($T_4$).

### vii. Limitations and recommendations

In the current study, we have delved into the influences of turbulent boundary layer flow on heat transport and flow mechanisms within the street canyon, leveraging high-resolution experimental data. However, it is important to acknowledge certain limitations in our work, along with open questions that beckon further investigation.

Firstly, controlling boundary layer flow conditions precisely in an experimental setting poses inherent challenges. To systematically differentiate the impacts of flow turbulence, it becomes essential to generate various inflows characterised by identical average velocities and turbulent boundary layer thickness but varying turbulence intensities. For instance, when examining the summarised plots of vertical velocity generated by different turbulent boundary layer flows, it proves challenging to discern significant differences. Furthermore, a more in-depth quantitative analysis of the thermal buoyant force is required to elucidate the precise effects of inflow turbulence on vertical buoyant updraft flow, a dimension that remains unclear.

While it is evident that stronger inflow fluctuations may suppress outflow and limit ventilation, at this juncture, we lack sufficient evidence to definitively distinguish between the impacts of mean and fluctuating components on ventilation rates. Hence, further investigation and analysis are imperative to shed light on this complex aspect. Similarly, we have presented the impacts of turbulence on heat removal capacity. However, a more comprehensive exploration involving flow scenarios with consistent average flow attributes but varying turbulence levels could provide invaluable insights into disentangling the respective contributions of convective and turbulent heat transfer within the context of non-isothermal flows.

Moving forward, numerical investigations stand to benefit from the detailed experimental data we have amassed. Upon successful validation, CFD simulations can hone in on specific questions, with the ability to isolate the impacts of other factors. These limitations, unresolved questions, and suggestions for future research avenues collectively contribute to the evolving discourse in this field.

### IV. Conclusion

In this experimental study, we investigate the impacts of turbulent boundary layer flow on heat and fluid flow within and over street canyons. The experiments are conducted in a closed-circuit water tunnel, and simultaneous

a) Corresponding author. Email: Yunpeng.xue@sec.ethz.ch

PIV-LIF techniques are employed to measure the velocity and temperature profiles. Three different turbulent boundary layer flows (TBLs) are created using flow roughness elements. TBL2 has a similar boundary layer thickness but slightly stronger turbulence than TBL1. TBL3 is a highly turbulent boundary layer flow with a lower inflow velocity at canyon height compared to TBL1 and TBL2.

We first test different ground temperatures and freestream velocities in the first turbulent boundary layer flow (TBL1) to gain a general understanding of the flow behaviour and temperature variation within and over the street canyons. The ground temperature, which determines the bulk Richardson number of the flow, showed significant impacts on the flow behaviour, particularly on the updraft buoyant flow and ventilation from the canyon opening.

By comparing the different parameters in three different TBLs, we found that when the models are submerged in the boundary layer flow, the flow and heat flux can be significantly influenced by the boundary layer profile. In our study with the selected TBLs, we observe a maximum difference of 80% in canyon ventilation between TBL3 and TBL1 when the freestream velocity and ground surface temperature are kept the same. The most significant difference in heat removal capacity, approximately 45%, is observed between TBL3 and TBL1, underscoring the crucial role of turbulent boundary layers (TBLs) in non-isothermal flows. Consequently, air ventilation and heat removal within the street canyon are enhanced when inflow of air into the street canyon is facilitated.

The fluctuating component of the flow is also significantly influenced by the turbulence intensity of the approaching boundary layer flow. TBLs with stronger fluctuation result in a larger region with strong fluctuation impacted by the thermal buoyant force. The integration of momentum flux ($u'v'$) reveals the evident influence of flow turbulence and the shear layer flow at the canyon opening. However, this influence is mainly significant around the canyon roof level and weaker at lower locations within the canyon.

The results reported in this work underscore the influence of turbulent boundary layer flow on heat and fluid flow characteristics in street canyons during non-isothermal flow conditions. This calls for further investigations into the turbulent characteristics of the boundary layer. An advanced understanding of the boundary layer is highly recommended for numerical simulations, particularly when dealing with smaller model scales or thick boundary layers, indicating that the flow region is within the boundary layer.

**Availability of data**
The data that support the findings of this study are available from the corresponding author upon reasonable request.

[a] Corresponding author. Email: Yunpeng.xue@sec.ethz.ch

a) Corresponding author. Email: Yunpeng.xue@sec.ethz.ch

[a] Corresponding author. Email: Yunpeng.xue@sec.ethz.ch

[a] Corresponding author. Email: Yunpeng.xue@sec.ethz.ch